\documentstyle[aps,aipbook,floats]{revtex}
\begin{document}

\vspace*{-1.0in}
\begin{flushright}
FERMILAB-Conf-95/307-T
\end{flushright}
\vspace*{0.25in}

\title{TOP QUARK PRODUCTION DYNAMICS
\footnote{
Invited talk at the X International $\bar{P}P$ Workshop,
Fermilab, Illinois, May 9 - 13, 1995.}
}

\author{Stephen Parke$^{\dagger}$}
\address{
$^{\dagger}$~parke@fnal.gov\\
Fermi National Accelerator Laboratory\thanks{
Fermilab is operated by the Universities Research Association under
contract with the United States Department of Energy.}
\\
Batavia, IL, 60510\\
USA}

\maketitle

\begin{abstract}
I review standard top quark production at the Fermilab Tevatron.
The current theoretical understanding of the total cross section and many
partial differential cross sections is presented.
Studies on the effects of extra gluon radiation on the top quark mass
determination are reviewed.
The possibility of new mechanisms for $t\bar{t}$ production are also
discussed.
\end{abstract}

\vspace{1.0cm}

\section*{Introduction}
Top quark production at the Fermilab Tevatron probes very high mass scales,
$ {\cal{O}}(500 ~GeV)$, and therefore is sensitive to new physics at this
scale.
Hence, it is important that we studied this process with high precision
and compare the results with the standard model predictions.
Also the top quark mass
along with the W-boson mass can be used as an indirect measurement
of the Higgs boson mass or even demonstrate the failure
of the standard model. Therefore we must measure the top quark mass
with the greatest precision possible with the events available to us.

Here I review the status of the cross section calculations for top
quark production not only for the total cross section but also for
the shape of the various distributions associated with the production.
For the kinematic measurement of the top quark mass the effects of
extra gluon radiation are important for precision measurements.
I will discuss
the status of our understanding of this extra radiation. Finally,
I will present a few of the possibilities for new physics that
can dramatically change top quark production not only in total
rate but also in the shape of the kinematic distributions.

\newpage
\section*{Cross Section}
\subsection*{Total}

In hadron colliders the dominant mode of top quark production is via
quark-antiquark annihilation or gluon-gluon fusion,
\begin{eqnarray*}
q~\bar{q} & \rightarrow & t ~\bar{t} \\
g~g  & \rightarrow & t ~\bar{t}.
\end{eqnarray*}
Currently, the most accurate determination of the QCD top cross section
is the Resummed Next to Leading Order (Re$\sum$ NLO) calculation of Laenen,
Smith and van Neerven\cite{lsvn}. Fig. 1 is a comparison of this Re$\sum$ NLO
calculation and the exact NLO calculation of Ellis\cite{rke} for both the
$q\bar{q}$ channel and the $gg$ channel at the Fermilab Tevatron.
Clearly, the $gg$ channel has
larger corrections from these resummed soft gluons.
However, for large top quark mass at the Tevatron, this
channel is only a small part of the total cross section.

\begin{figure}[hbt]
\vspace{8cm}
\includegraphics{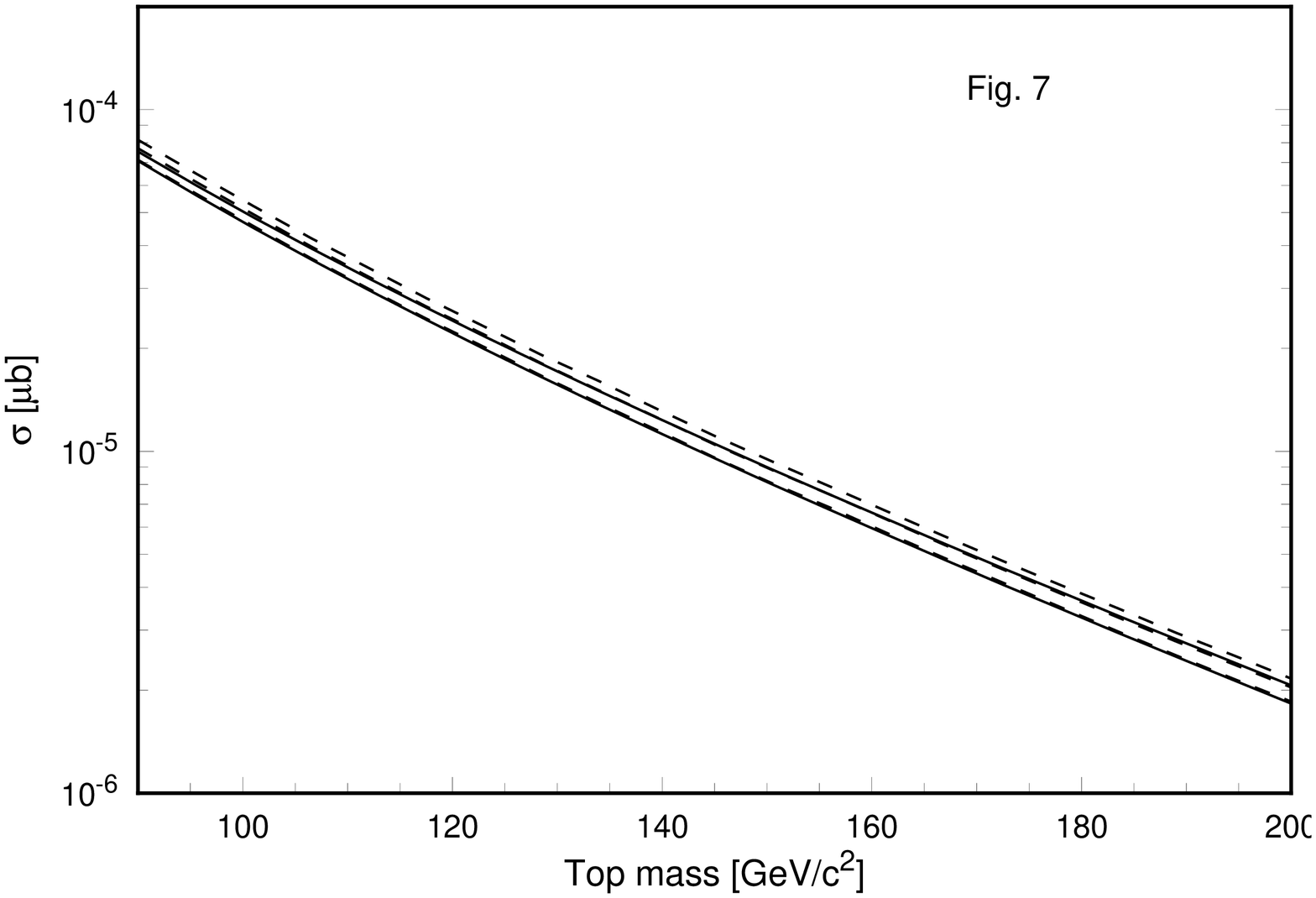}
\includegraphics{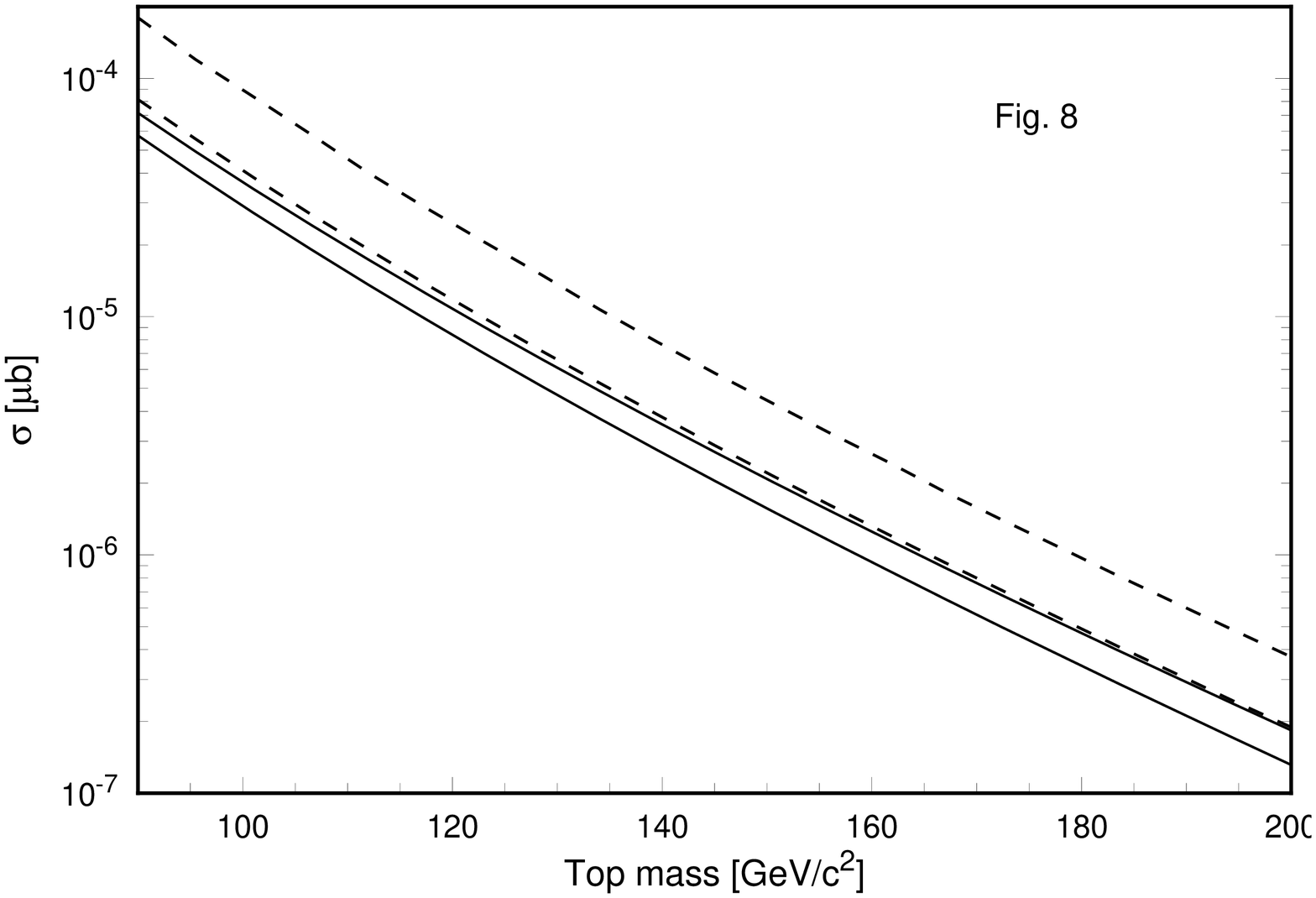}
\vspace{-1.25cm}\hspace{1.0cm}(a)\hspace{5.75cm}(b)\vspace{1cm}
\vspace{-0.5cm}
\caption[]{Range of cross sections for the exact NLO calculation (solid) and
the Resummed NLO calculation (dashed): (a) for the $q\bar{q}$ channel only,
(b) for the $gg$ channel only.}
\label{channels}
\end{figure}

Fig. 2 is a
similar comparison for the total cross section.
At large top quark masses the
difference between the Re$\sum$ NLO calculation and the NLO calculation
is at the 20\% level. At the LHC where the $gg$ channel will be the dominant
production mode these resummed corrections will be a much larger correction
to the total cross section.

\begin{figure}[t]
\vspace{8cm}
\includegraphics{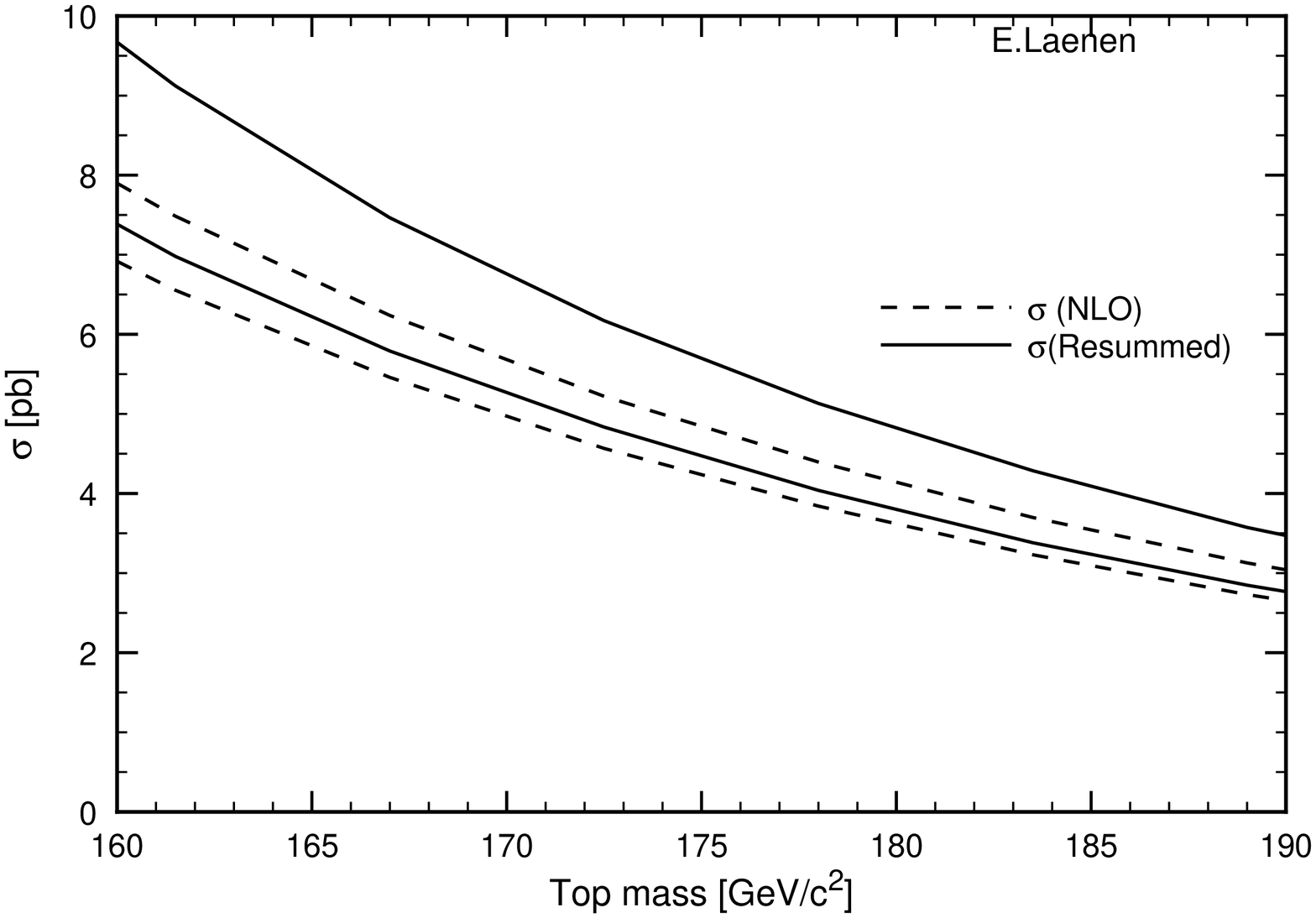}
\vspace{0cm}
\caption[]{The total QCD Top Quark Production Cross
Section at the Tevatron.
The solid curves give the range of values
using the resummed next to leading order
calculations and the dashed curves are the range
for the next to leading order calculations.
The same structure functions were used.}
\label{xsectotal}
\end{figure}

The resummation technique of Laenen, Smith and van Neerven
involves a infra-red cutoff which is set appropriately.
Contopanagos and Sterman \cite{cs}
have pioneered a resummation technique, principal value resummation,
which does not involve such an arbitrary cutoff. This method has been applied
to top production at the Tevatron by Berger and Contopanagos \cite{bc}.
At the time of this conference they had completed the $q\bar{q}$ channel
but were still working on the $gg$ channel.
The results using the principal value resummation for
the less sensitive $q\bar{q}$ channel are in close agreement with
the infra-red cutoff resummation. The completion of this calculation
for the total cross section is important for
determining how well we know the total top quark cross section.

\subsection*{Shapes}

\begin{figure}[t]
\vspace{6cm}
\includegraphics{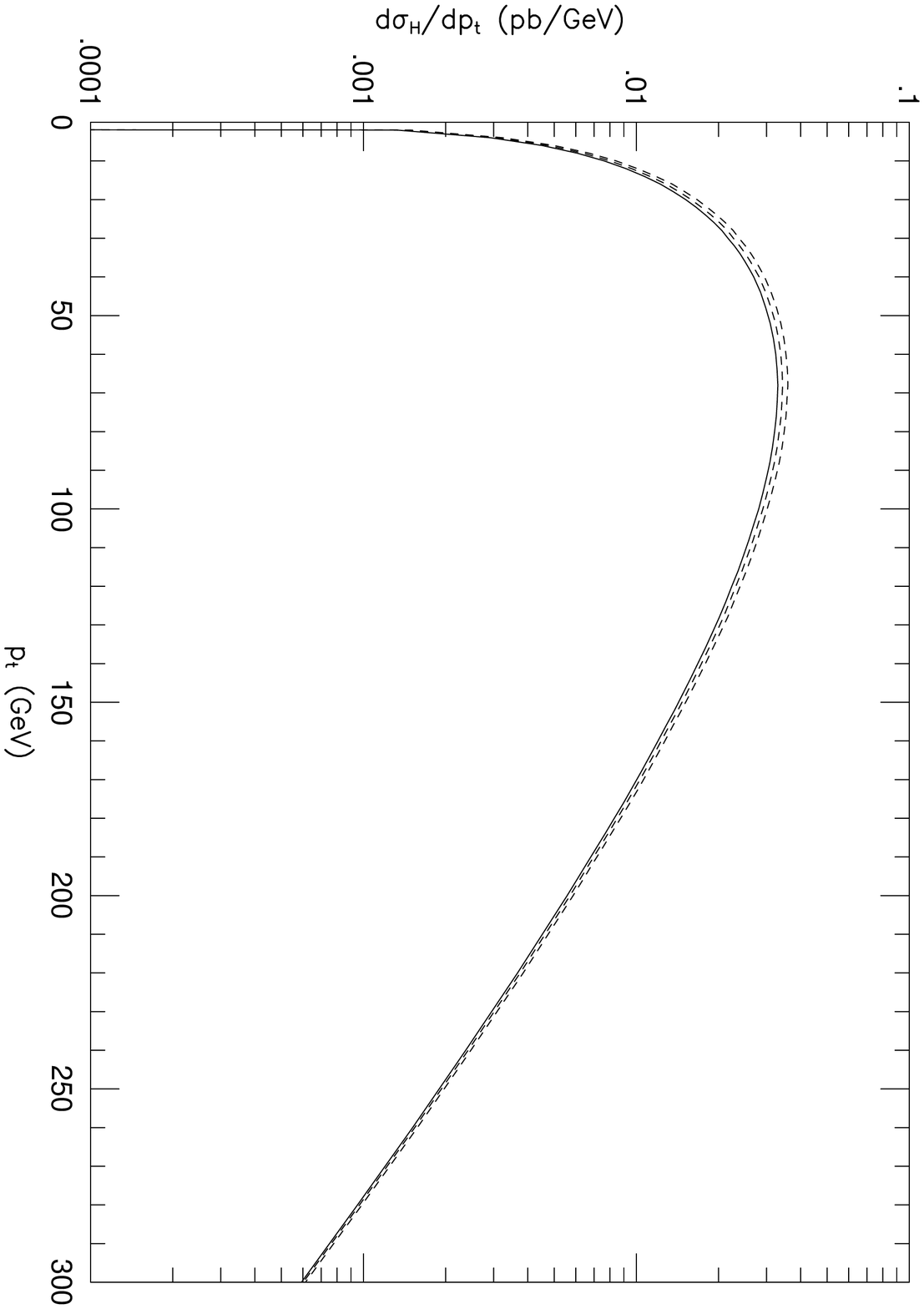}
\includegraphics{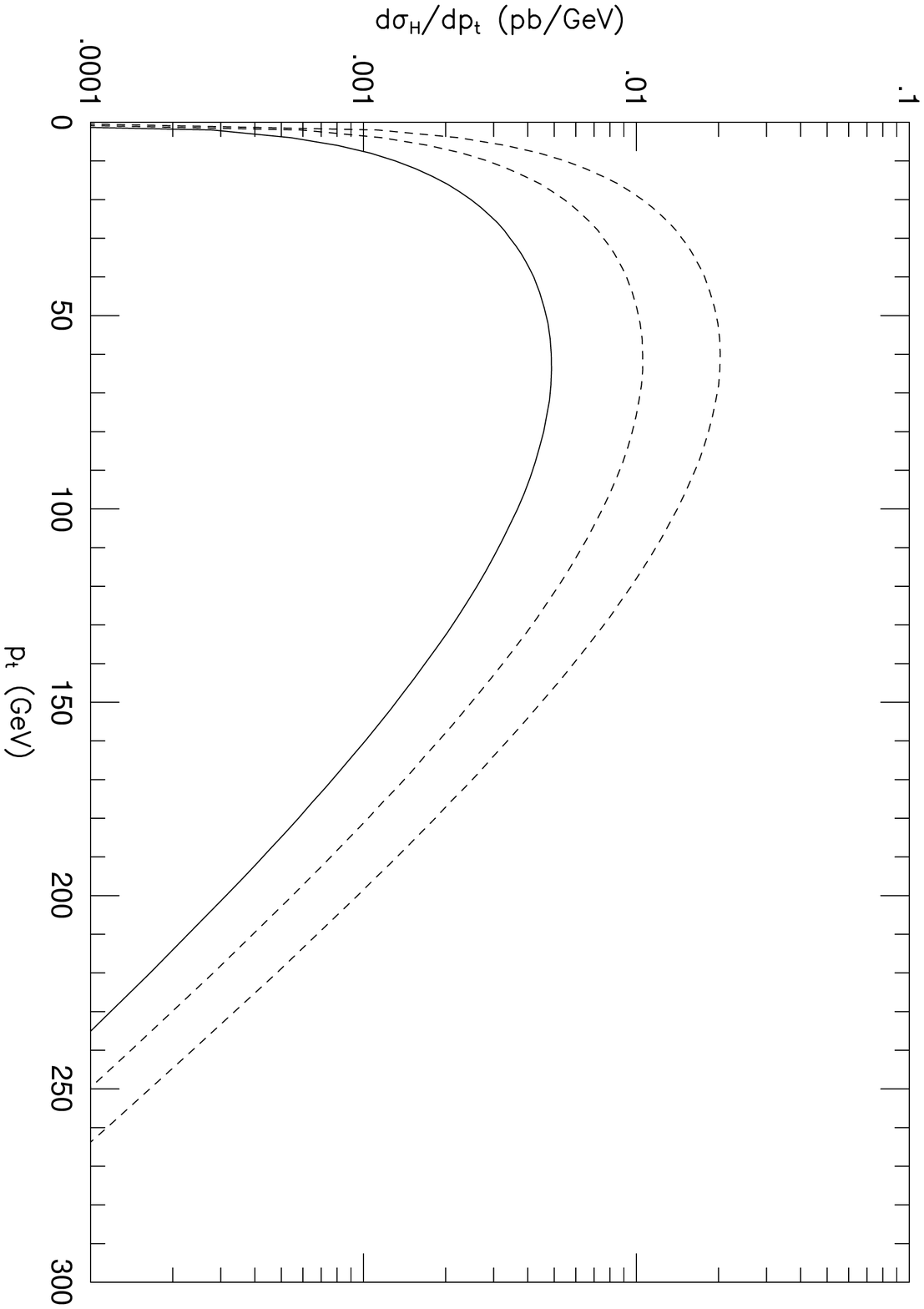}
\includegraphics{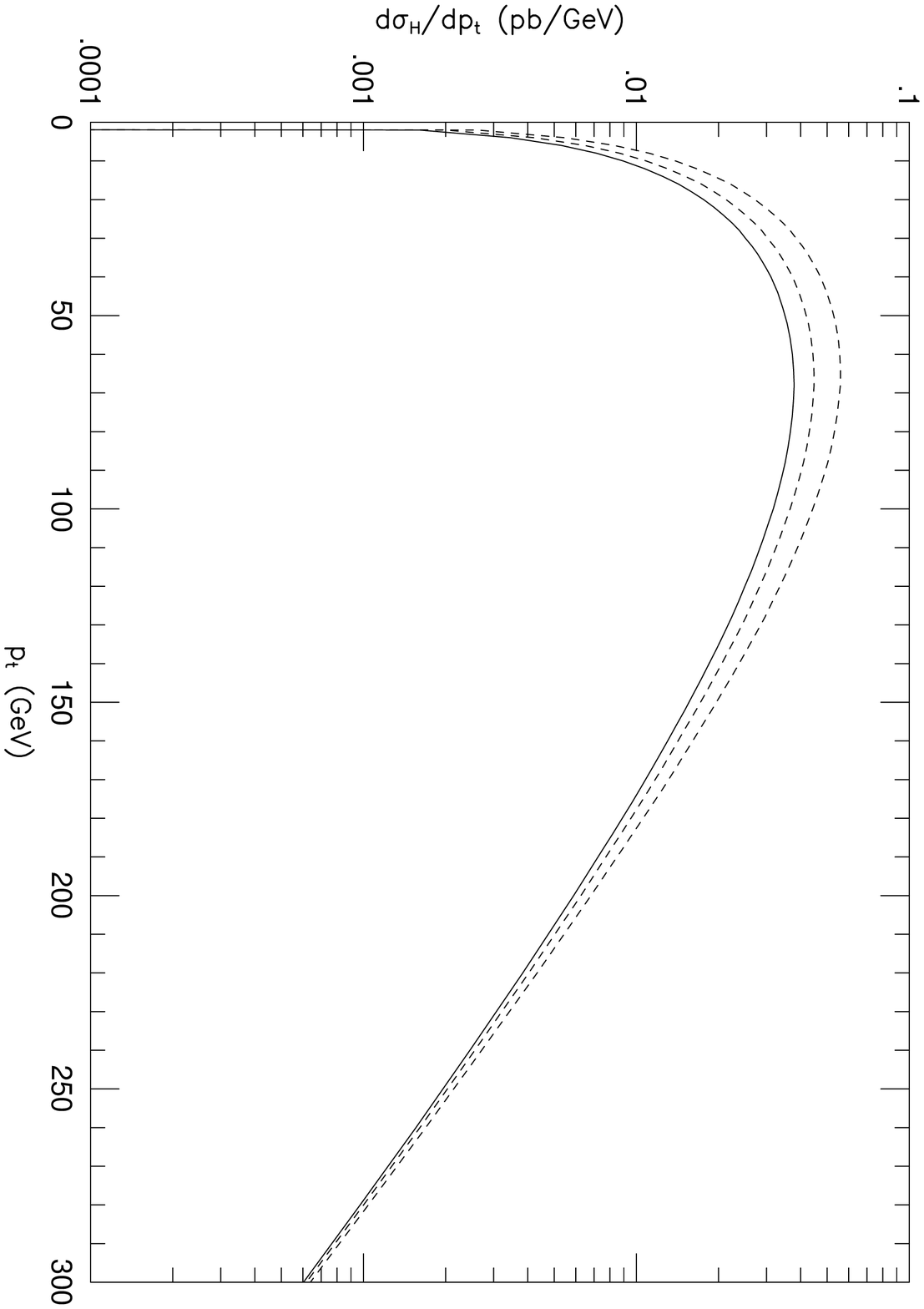}
\vspace{-1.25cm}\hspace{1.0cm}(a)\hspace{3.5cm}(b)\hspace{3.5cm}(c)\vspace{1cm}
\vspace{-0.75cm}
\caption[]{
The top quark $p_t$ distribution for 175 GeV top quark mass.
Solid line is NLO and the dashed lines are the range of values for
the Re$\sum$ NLO calculation.
For (a) $q\bar{q}$ channel, (b) $gg$ channel, (c) Sum.
}
\label{shapeksp}
\end{figure}

\begin{figure}[h]
\vspace{6cm}
\includegraphics{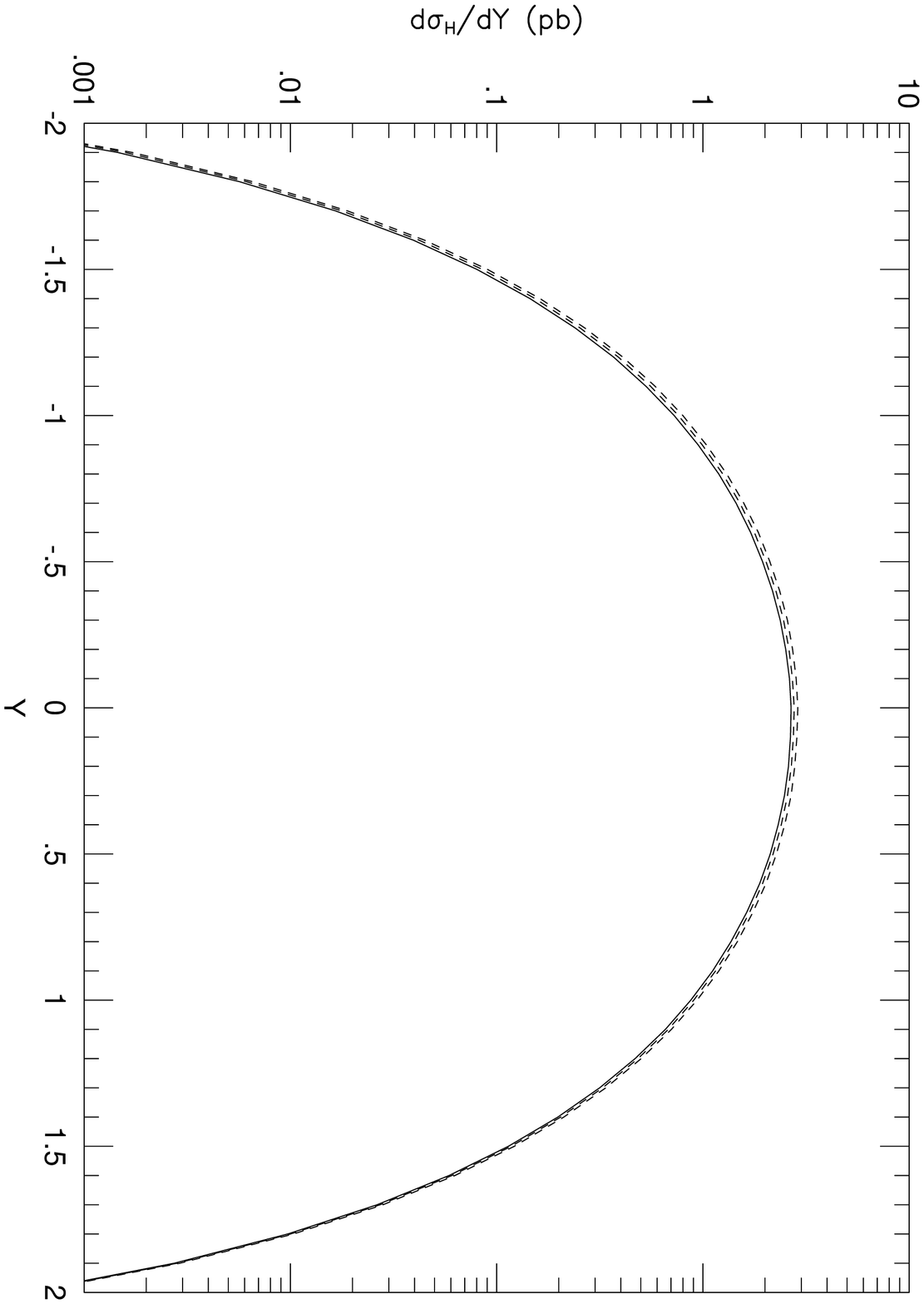}
\includegraphics{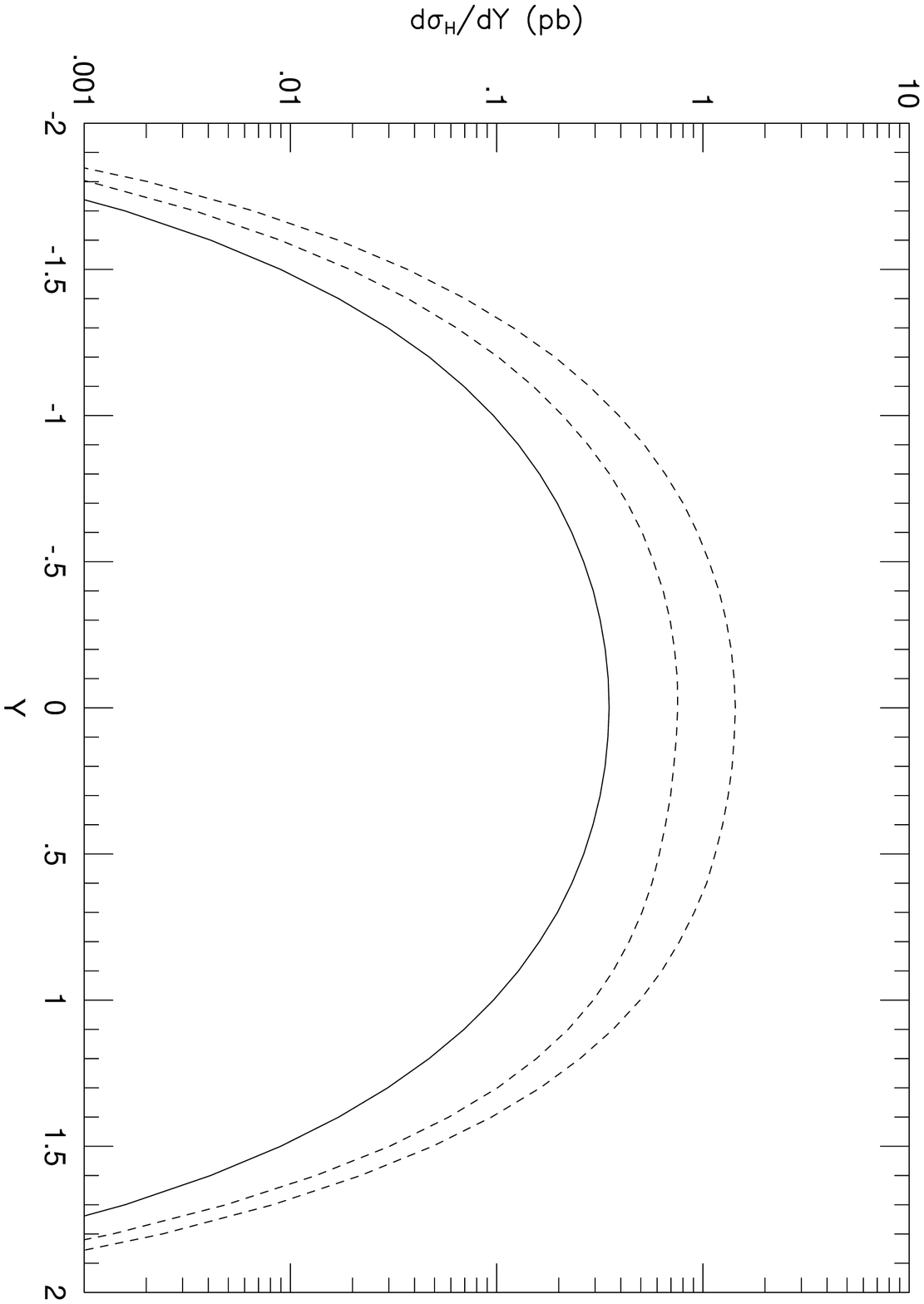}
\includegraphics{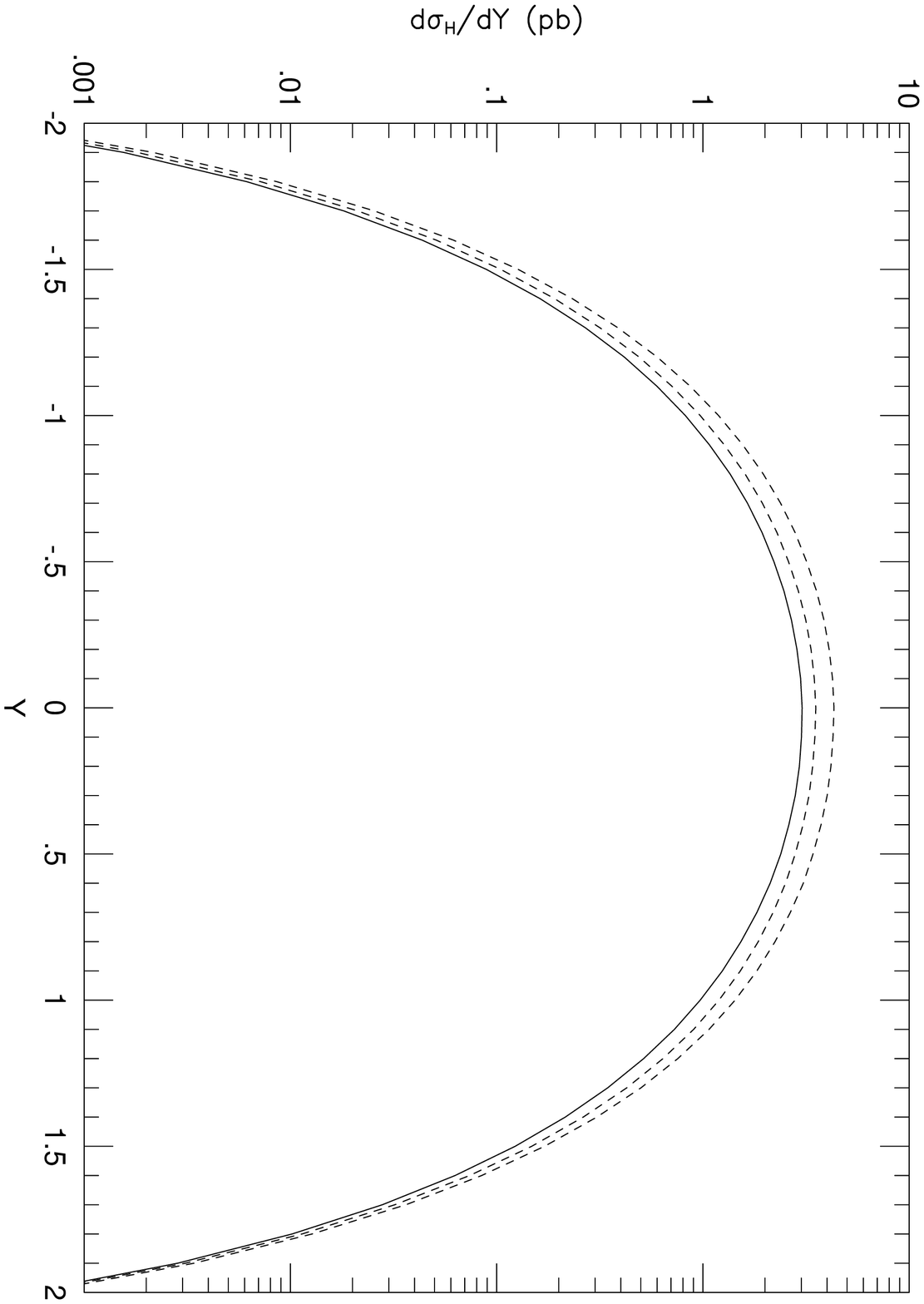}
\vspace{-1.25cm}\hspace{1.0cm}(a)\hspace{3.5cm}(b)\hspace{3.5cm}(c)\vspace{1cm}
\vspace{-0.75cm}
\caption[]{
The top quark rapidity distribution for 175 GeV top quark mass.
Solid line is NLO and the dashed lines are the range of values for
the Re$\sum$ NLO calculation.
For (a) $q\bar{q}$ channel, (b) $gg$ channel, (c) Sum.
}
\label{shapeksy}
\end{figure}

Kidonakis and Smith\cite{ks} have calculated the
inclusive transverse momentum and rapidity distributions for
top quark production at the Fermilab Tevatron using the infra-red
cutoff resummation technique.
A comparison of these results to the next leading order calculation
for the transverse momentum of the top quark is shown in Fig. 3 and for the
rapidity in Fig. 4.

\clearpage

Frixione, Mangano, Nason and Ridolfi \cite{fmnr} have calculated a number of
distributions for top quark production in NLO and made comparisons
with the HERWIG monte carlo. In Fig. 5, I show from their paper
the invariant mass distribution
of the $t\bar{t}$ pair, the transverse momentum distribution of the
$t\bar{t}$ pair and the azimuthal distribution of the $t\bar{t}$ pair.
These last two distributions are trivial at lowest order so it is satisfying
to see that the two calculations are comparable in the
region where we expect approximate agreement.

\begin{figure}[hbt]
\vspace{7cm}
\includegraphics{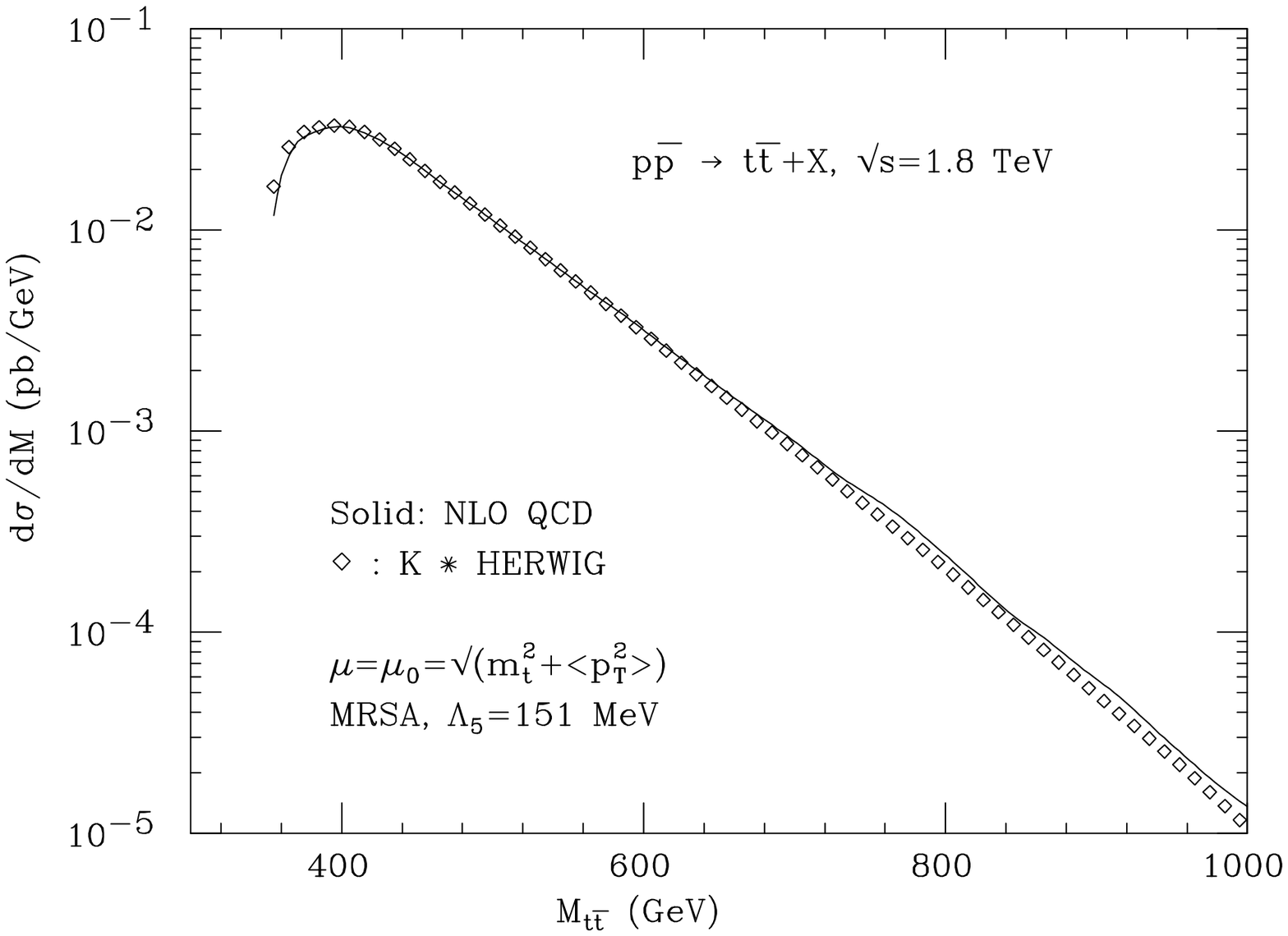}
\includegraphics{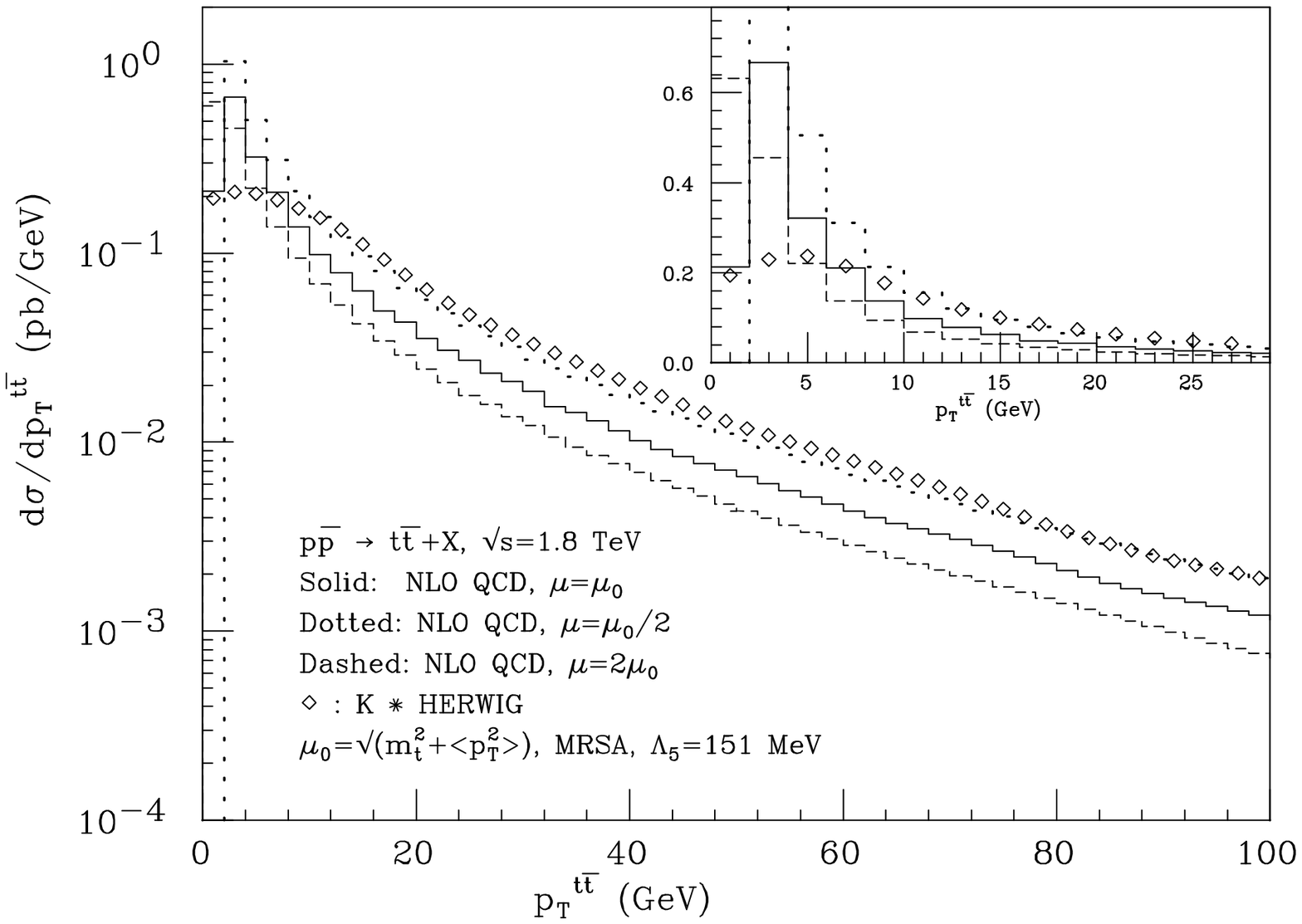}
\includegraphics{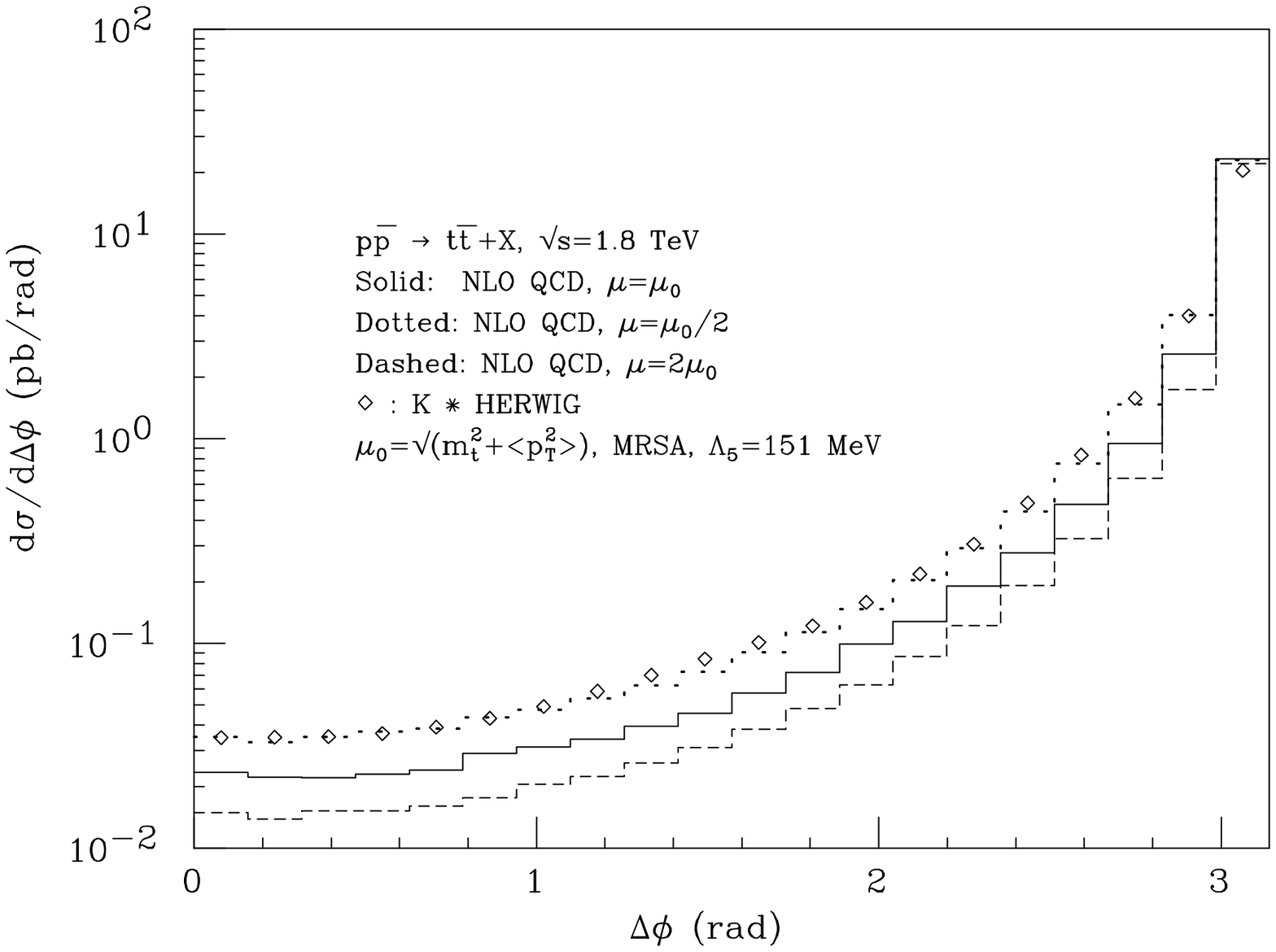}
\vspace{-1.25cm}\hspace{1.0cm}(a)\hspace{3.5cm}(b)\hspace{3.5cm}(c)\vspace{1cm}
\vspace{-0.75cm}
\caption[]{For a top quark mass of 174 GeV at NLO verses HERWIG for
(a) the invariant mass distribution of the $t\bar{t}$ pair,
(b) the transverse momentum distribution of the $t\bar{t}$ pair
and
(c) the azimuthal distribution of the $t\bar{t}$ pair.
}
\label{shapemlm}
\end{figure}

\section*{Extra Gluon Radiation}

For precision measurements of the top quark mass,
we need to understand the effects of extra gluon radiation in
top quark production. This has been studied by Lampe\cite{bl},
Orr and Stirling\cite{os} and most recently by
Orr, Stelzer and Stirling \cite{ossa}. In Fig. 6, I show the
results of Orr, Stelzer and Stirling, for the invariant mass of
the W-boson  and b-quark jet without and with an extra gluon jet.
Clearly the results of the mass fitting are sensitive to how this
extra jet is treated. Therefore it is important that we understand
this process very well for precision top quark mass measurements.

Orr, Stelzer and Stirling have also compared their exact tree-level
calculation with the HERWIG monte carlo, see Fig. 7. The difference
between these two calculation, I believe, is dependent 	  upon
how the top quark mass is included in the monte carlo.
This discrepancy is still to be resolved.
Further discussion on this
problem can be found in Orr, Stelzer and Stirling \cite{ossb} which
addresses this issue in the simpler environment of an e+e- collider.

\newpage

\begin{figure}[t]
\vspace{5cm}
\includegraphics{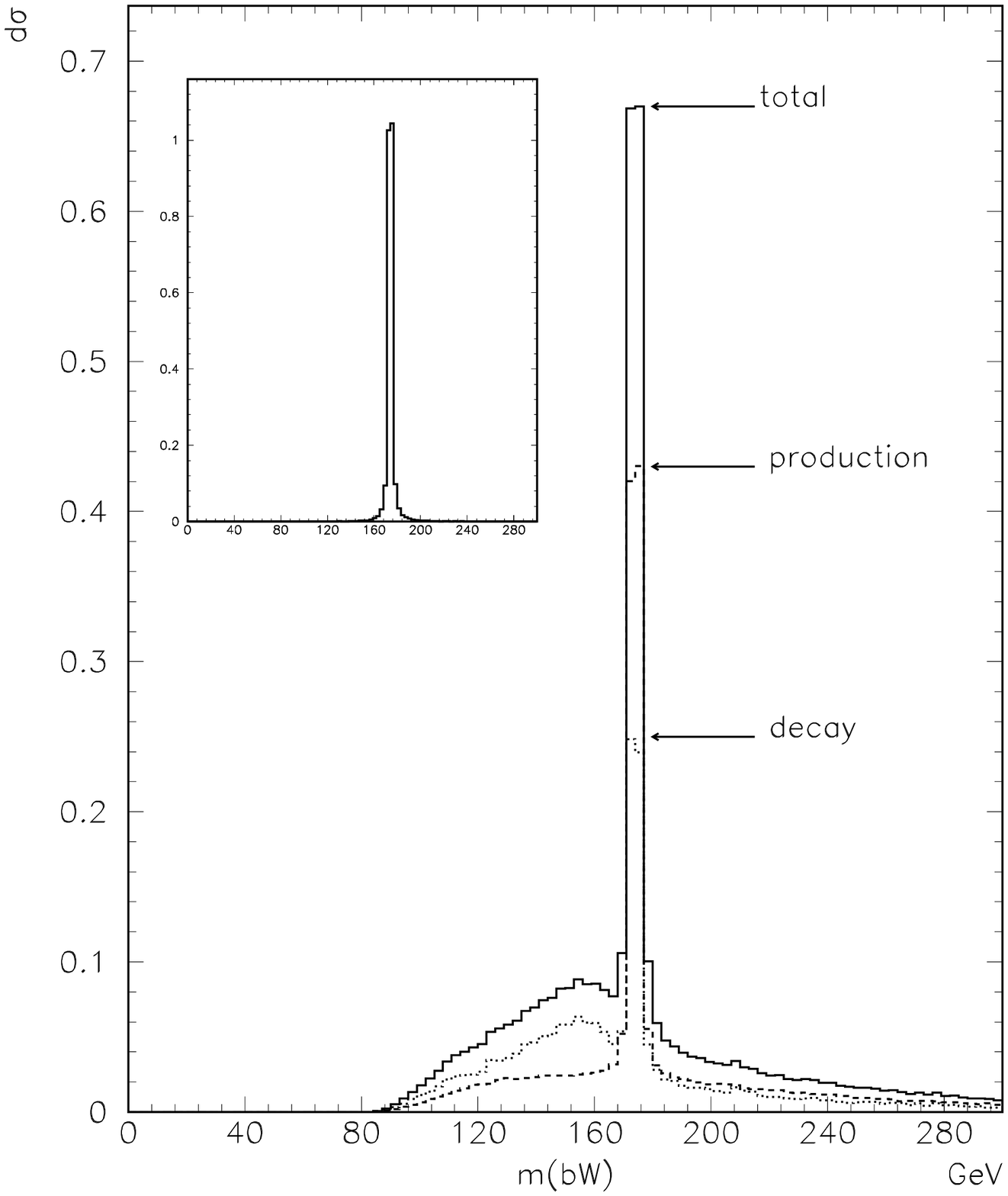}
\includegraphics{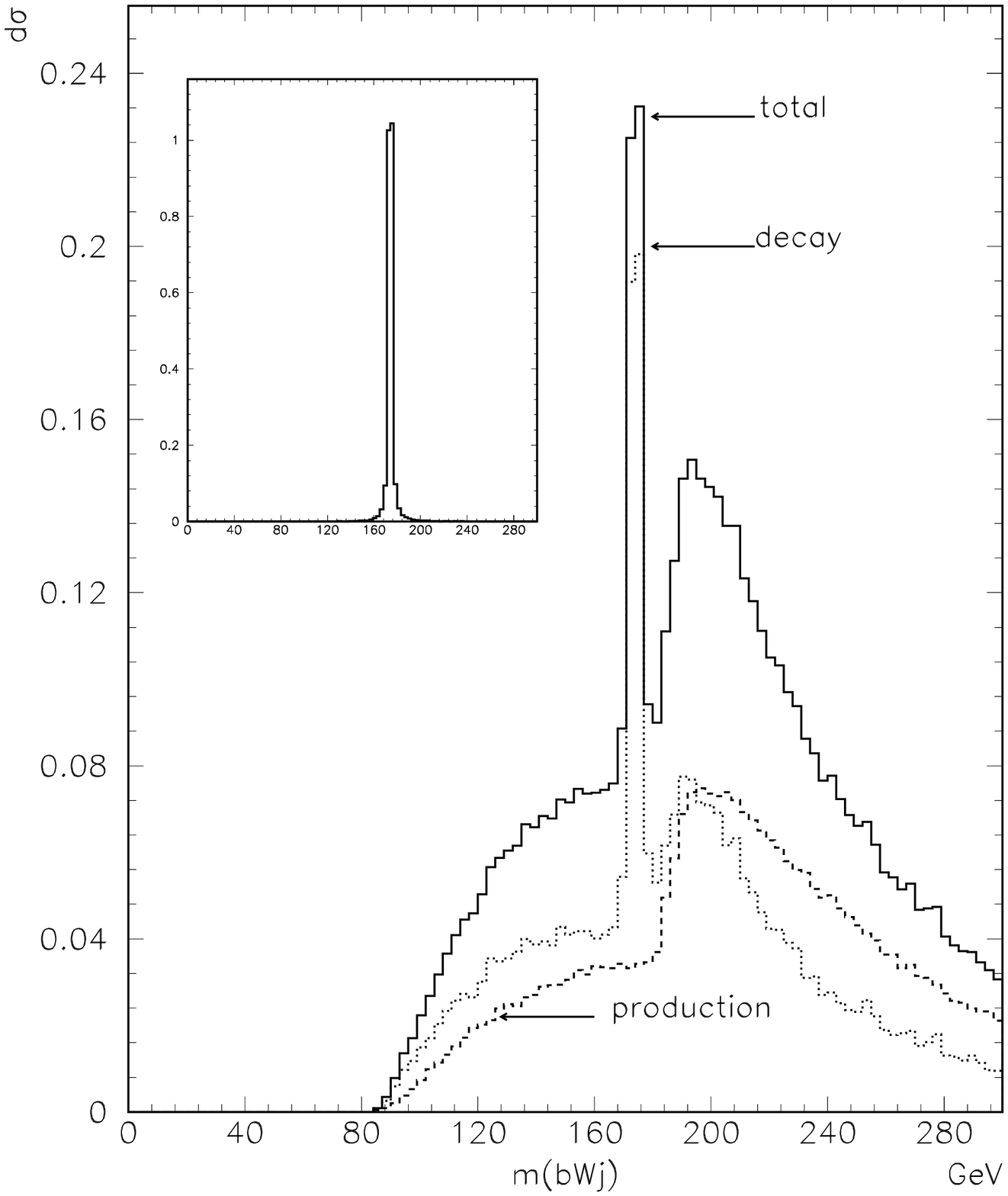}
\vspace{-1.25cm}\hspace{2.0cm}(a)\hspace{4.50cm}(b)\vspace{1cm}
\vspace{-1cm}
\caption[]{
(a) The distribution for the Wb invariant mass.
Also shown are the distributions corresponding to the production
(dot-dashed) and decay (dotted) emission contributions.
(b) The distribution for the Wb+jet invariant mass.
Also shown are the distributions corresponding to the production
(dot-dashed) and decay (dotted) emission contributions.
}\label{wbmass}
\end{figure}

\begin{figure}[b]
\vspace{12.5cm}
\includegraphics{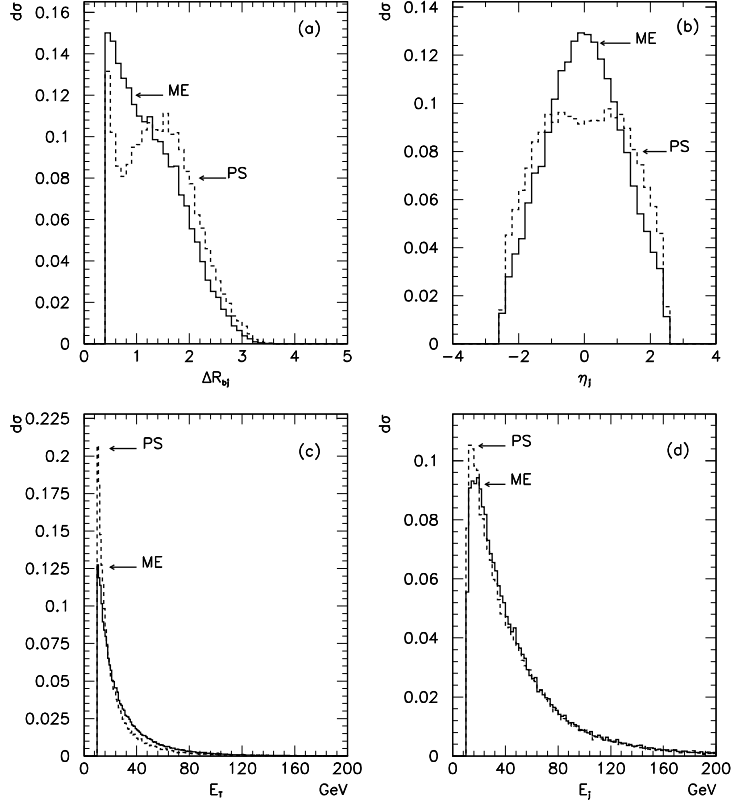}
\vspace{-1.5cm}
\caption[]{Distributions in (a) the jet-b separation $\Delta R_{bj}$,
(b) the jet pseudorapidity $\eta_j$,
(c) the jet $E_T$, and
(d) the jet energy $E_j$ in the subprocess center-of-mass frame,
for the exact calculation (solid, labeled ME)
and as obtained using the HERWIG parton-shower monte carlo program
(dashed, labeled PS).
}\label{herwig}
\end{figure}

\clearpage


\section*{New Dynamics}
Since the top quark mass is close to the electro-weak symmetry
breaking scale it is possible that top quark production
will provide an exciting window on new physics.

\begin{figure}[hbt]
\vspace{7cm}
\includegraphics{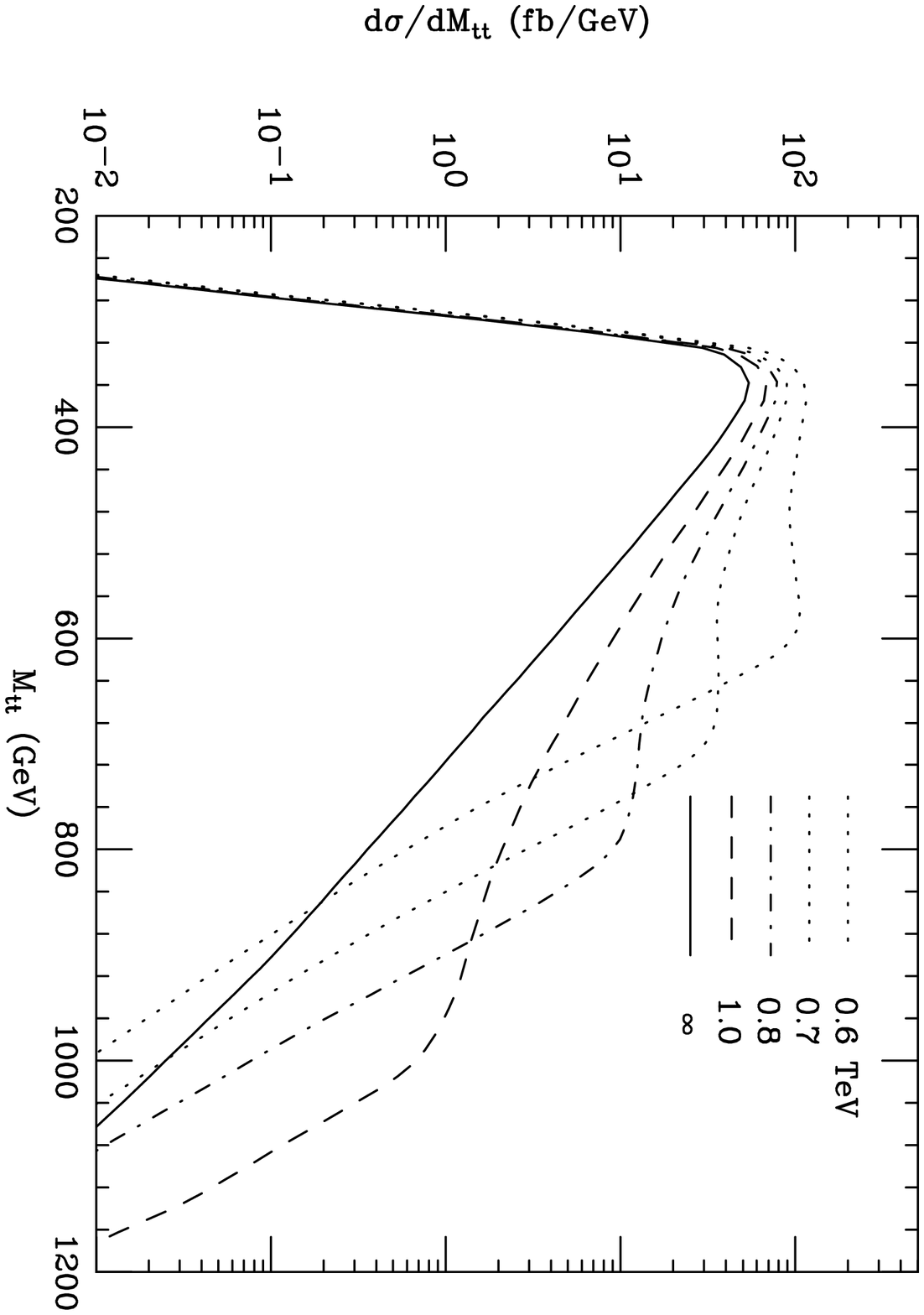}
\includegraphics{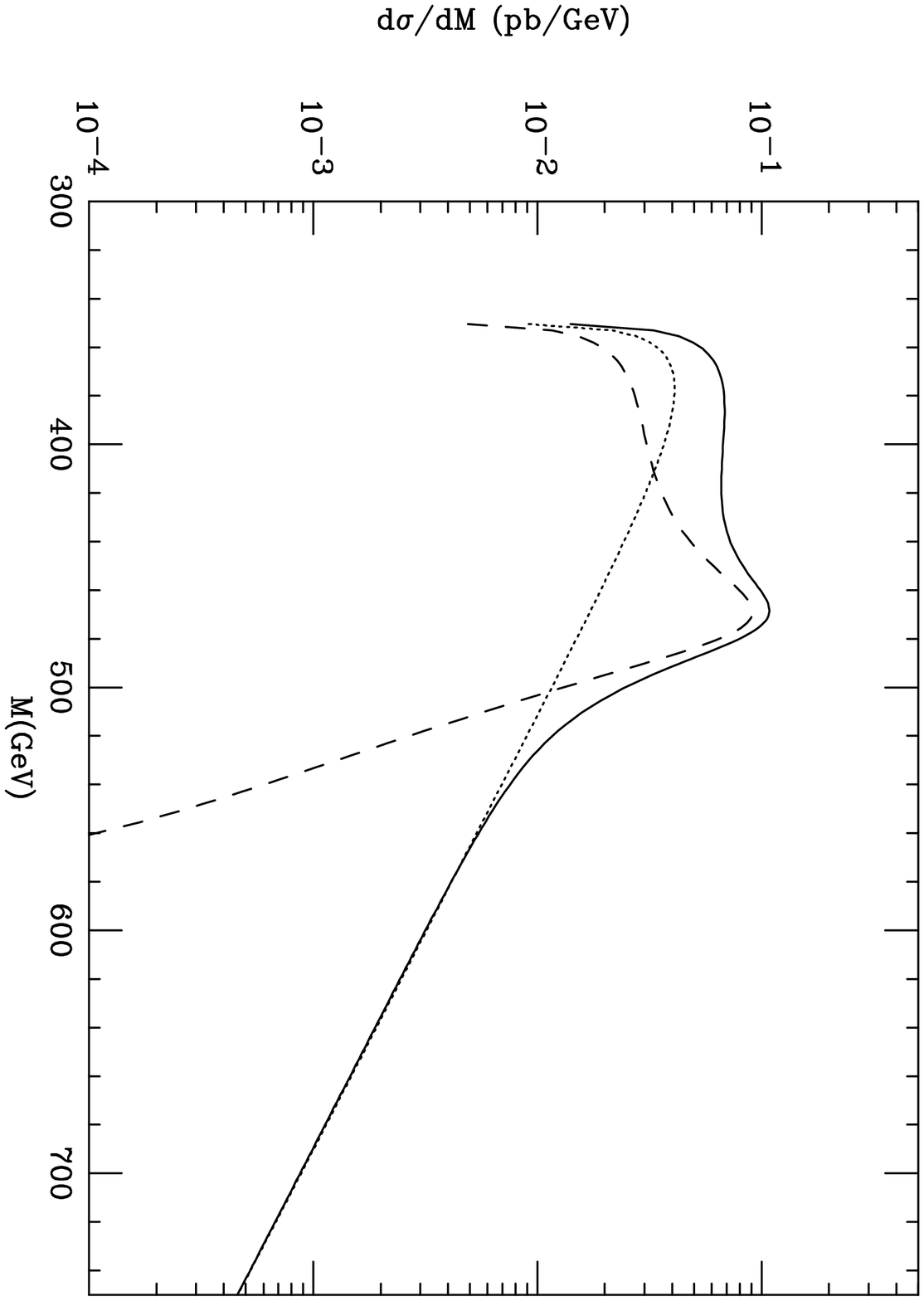}
\vspace{-1.25cm}\hspace{1.0cm}(a)\hspace{5.75cm}(b)\vspace{1cm}
\vspace{-0.5cm}
\caption[]{The invariant mass of the $t\bar{t}$ pair for (a)
topcolor octet model and (b) the two scale technicolor model.}
\label{newdyn}
\end{figure}

Hill and Parke
\cite{hp} have shown the effects of new physics on top
production in the $q\bar{q}$ channel using both a general
effective Lagrangian approach as well as in a specific top
color model. Fig. 8(a) gives the distortions in the $t\bar{t}$
invariant mass in the color octet top color model for various masses
of the top color boson.

Later Eichten and Lane \cite{el} discussed the effects on top quark
product in a multi-scalar technicolor model. Here the enhanced
production is in the $gg$ channel. Fig. 8(b) is their invariant mass
plot.

\section*{Conclusions}
The total cross section for $t\bar{t}$ production is in good shape
for the Fermilab Tevatron. New calculations which include higher
order effects give small contributions. Therefore the uncertainties
in this cross section are well under control. Many differential
cross section have also been calculated, some simple ones
using the Re$\sum$ NLO techniques while others which are trivial at
tree level have been calculated at NLO.

The effects of extra gluon radiation in the determination of the
top quark mass is still under study. I expect that
the discrepancy between the
tree level matrix element calculation and the HERWIG monte carlo
will be resolved in the near future.

The possibility of finding new physics in $t\bar{t}$ production is
very exciting. Kinematic distributions for top production and
decay are eagerly waited. Watch out for surprises!

I wish to thank all the authors of the references
who help me in the preparation of this presentation.

\end{document}